\documentclass[aps,pre,showpacs,noshowkeys,amsmath,amssymb,amsfonts,superscriptaddress,longbibliography,reprint]{revtex4-1}
\usepackage[english]{babel}

\usepackage{graphicx}
\usepackage{bm}
\usepackage{physics}
\usepackage{mathtools}
\usepackage{gensymb}
\setcitestyle{super}
\usepackage{caption}
\usepackage{subcaption}
\DeclareCaptionLabelSeparator{bar}{~\rule[-0.4ex]{0.2ex}{1em}~}
\DeclareCaptionLabelFormat{subfor}{\textbf{#2}}
\newcommand*\bfcaption[2]{\caption[#1]{\textbf{#1.}#2}}
\usepackage{xcolor}
\definecolor{UBcolor}{HTML}{007CC1}
\usepackage[colorlinks=true,pdfnewwindow=true,linkcolor=UBcolor,citecolor=UBcolor,urlcolor=UBcolor,breaklinks=true,linktocpage]{hyperref}
\usepackage[all]{hypcap}
\usepackage[nameinlink,capitalise]{cleveref}
\begin{document}

\title{Topological defects promote layer formation in \textit{Myxococcus xanthus} colonies}

\author{Katherine Copenhagen}
\altaffiliation{These authors contributed equally to this work.}
\affiliation{Lewis-Sigler Institute for Integrative Genomics, Princeton University, Princeton, NJ 08544, USA}

\author{Ricard Alert}
\altaffiliation{These authors contributed equally to this work.}
\affiliation{Lewis-Sigler Institute for Integrative Genomics, Princeton University, Princeton, NJ 08544, USA}
\affiliation{Princeton Center for Theoretical Science, Princeton University, Princeton, NJ 08544, USA}

\author{Ned S. Wingreen}
\affiliation{Lewis-Sigler Institute for Integrative Genomics, Princeton University, Princeton, NJ 08544, USA}
\affiliation{Department of Molecular Biology, Princeton University, Princeton, NJ 08544, USA}

\author{Joshua W. Shaevitz}
\affiliation{Lewis-Sigler Institute for Integrative Genomics, Princeton University, Princeton, NJ 08544, USA}
\affiliation{Joseph Henry Laboratories of Physics, Princeton University, Princeton, NJ 08544, USA}
\date{\today}

\maketitle

\textbf{The soil bacterium \textit{Myxococcus xanthus} lives in densely packed groups that form dynamic three-dimensional patterns in response to environmental changes, such as droplet-like fruiting bodies during starvation \cite{Kaiser2003}. The development of these multicellular structures begins with the sequential formation of cell layers in a process that is poorly understood \cite{Curtis2007}. Using confocal three-dimensional imaging, we find that motile rod-shaped \textit{M. xanthus} cells are densely packed and aligned in each layer, forming an active nematic liquid crystal. Cell alignment is nearly perfect throughout the population except at point defects that carry half-integer topological charge. We observe that new cell layers preferentially form at the position of $+1/2$ defects, whereas holes preferentially open at $-1/2$ defects. To explain these findings, we model the bacterial colony as an extensile active nematic fluid with anisotropic friction. In agreement with our experimental measurements, this model predicts an influx of cells toward $+1/2$ defects, and an outflux of cells from $-1/2$ defects. Our results suggest that cell motility and mechanical cell-cell interactions are sufficient to promote the formation of cell layers at topological defects, thereby seeding fruiting bodies in colonies of \textit{M. xanthus}.}

The rod-shaped soil bacterium \textit{Myxococcus xanthus} lives in colonies of millions of individual cells that migrate on surfaces. These colonies exhibit a wide range of motility-driven collective behaviors including rippling during predation and fruiting body formation in response to starvation \cite{Kaiser2003,Zhang2012,Kim1990,Jelsbak2002}. When nutrients are scarce, individual cells alter their motility to drive the population from a thin sheet coating the underlying substrate to a series of dome-shaped multicellular structures called fruiting bodies \cite{Liu2019a}. The first step in this process is the sequential formation of cell layers on top of an original cell monolayer \cite{Curtis2007,Kaiser2014}. However, the physical mechanism underlying layer formation remains largely unknown, partly because the rapid development of fruiting bodies makes it difficult to monitor the emergence of new layers in detail.

Here, to overcome this limitation and address how new cell layers emerge from preexisting ones, we placed \textit{M. xanthus} cells on an agar substrate in the presence of nutrients. In these conditions, the dimensionless inverse P\'{e}clet number that characterizes the persistence of cell migration is $\mathrm{Pe}_{\text{r}}^{-1} = 1.28\pm 0.09$ (\hyperref[methods]{Methods}, \cref{motility}). At this value of $\mathrm{Pe}_{\text{r}}^{-1}$, with an average cell density of $\rho_0 = 0.25\pm 0.04$ cells/$\mu$m$^2$ (S.D.), the colony does not form fruiting bodies but remains as a thin sheet wetting the substrate \cite{Liu2019a}. Nevertheless, new cell layers and holes spontaneously appear and disappear (\hyperref[movies]{Movies S1 and S2}), allowing us to examine these processes in detail.

\begin{figure}[tb]
\begin{center}
    \includegraphics[width=\columnwidth]{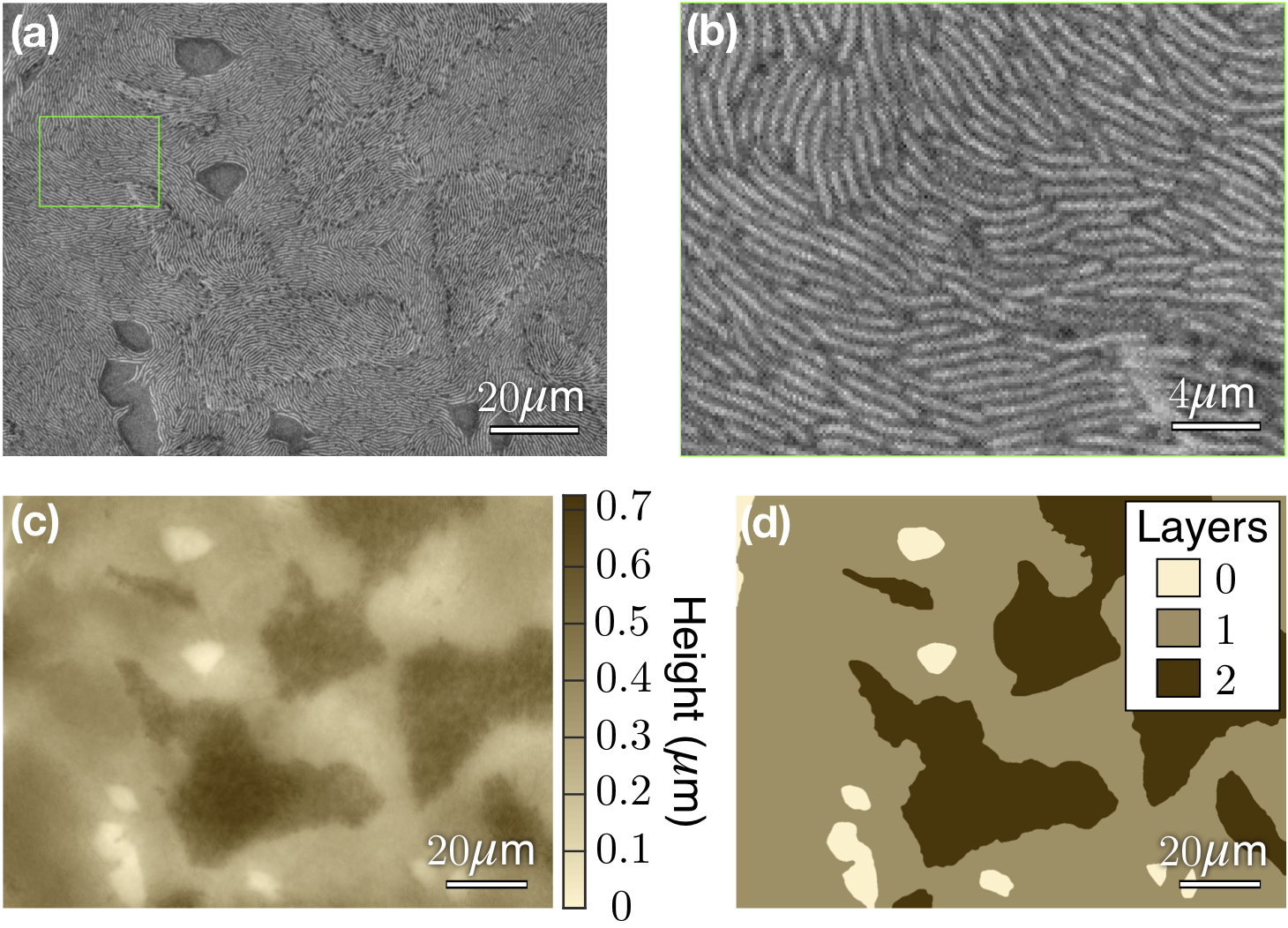}
\end{center}
  {\phantomsubcaption\label{brightness}}
  {\phantomsubcaption\label{brightness-closeup}}
  {\phantomsubcaption\label{height}}
  {\phantomsubcaption\label{layers}}
    \bfcaption{Dense colonies of \textit{M. xanthus} form layered active nematic liquid crystals}{ Sub-cellular resolution images of the reflectance field (\subref*{brightness}, \subref*{brightness-closeup}), height field (\subref*{height}), and layer-count field (\subref*{layers}) of a layered \textit{M. xanthus} colony. $0$ layers corresponds to the height of the agar substrate. The zoomed-in view (\subref*{brightness-closeup}) of the green box in \subref*{brightness} reveals cell alignment.}
\end{figure}

To this end, we imaged the colony using a three-di\-men\-sio\-nal, laser scanning confocal microscope with subcellular resolution (\hyperref[methods]{Methods}). This instrument measures light reflected from the surface of the sample and does not require fluorescence or cell labeling. By sampling a stack of positions along the microscope axis, we simultaneously measured the height and the reflectance fields (\cref{brightness,brightness-closeup,height}). Images of the reflectance field revealed that the rod-shaped \textit{M. xanthus} cells are densely packed, aligned with neighboring cells, and retain their motility (\cref{brightness-closeup}, \hyperref[movies]{Movie S1}), driven by both cell-substrate and cell-cell interactions. The height field showed that the colony organizes into discrete layers rather than a continuous distribution of heights (\cref{height}). We thresholded the height data to measure the number of layers at every position in the image (\cref{layers}). Colonies of $\Delta$\textit{pilA} mutant cells lacking pili exhibit very similar behaviors (\hyperref[movies]{Movie S3}), showing that these extracellular appendages are not required for layer formation or hole opening.

\begin{figure}[t]
    \includegraphics[width=\columnwidth]{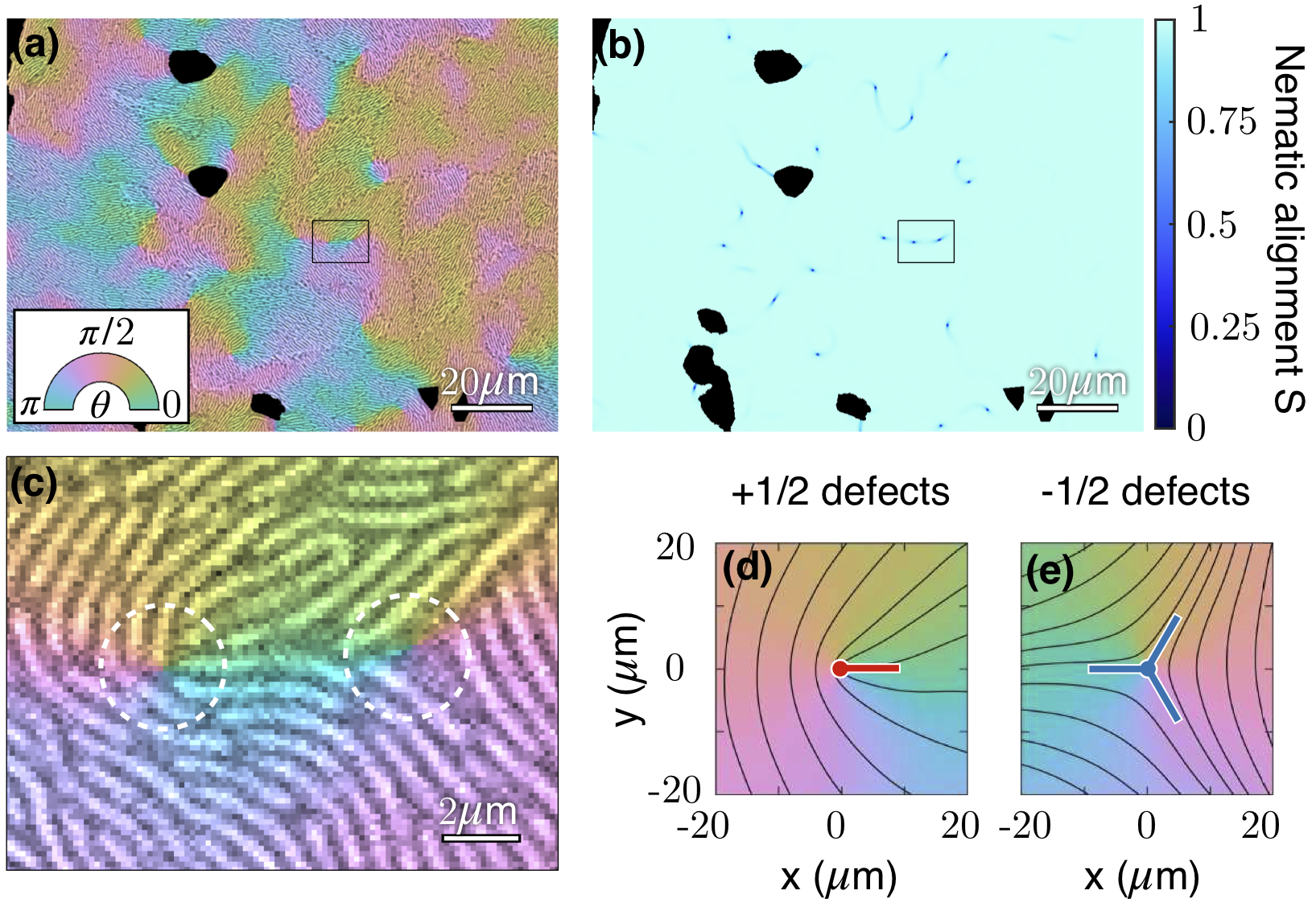}
  {\phantomsubcaption\label{angle-field}}
  {\phantomsubcaption\label{strength-field}}
  {\phantomsubcaption\label{angle-closeup}}
  {\phantomsubcaption\label{angle+1/2}}
  {\phantomsubcaption\label{angle-1/2}}
    \bfcaption{Nematic alignment and topological defects in \textit{M. xanthus} colonies}{ (\subref*{angle-field}) Nematic cell-orientation angle field $\theta(\bm{r})$ overlaid on a reflectance image of a colony. (\subref*{strength-field}) Nematic alignment field $S(\bm{r})$ for the region shown in \subref*{angle-field}. The points at which $S$ approaches $0$ are topological defects. (\subref*{angle-closeup}) Zoomed-in view of the black box in \subref*{angle-field}; the white dashed lines encircle a $+1/2$ defect (left) and a $-1/2$ defect (right) of the nematic cell alignment. (\subref*{angle+1/2}, \subref*{angle-1/2}) Average angle field measured around $+1/2$ and $-1/2$ defects, respectively. Lines indicate the local axis of cell alignment. $+1/2$ ($-1/2$) defects are labeled with a red (blue) point at the defect core and segments along their axes of symmetry. Averages are over a total of 7896 frames from 96 tracks of $+1/2$ defects and 7096 frames from 144 tracks of $-1/2$ defects, across 8 replicate experiments (\hyperref[methods]{Methods}).}
\end{figure}

\begin{figure*}[bt]
    \includegraphics[width=0.75\textwidth]{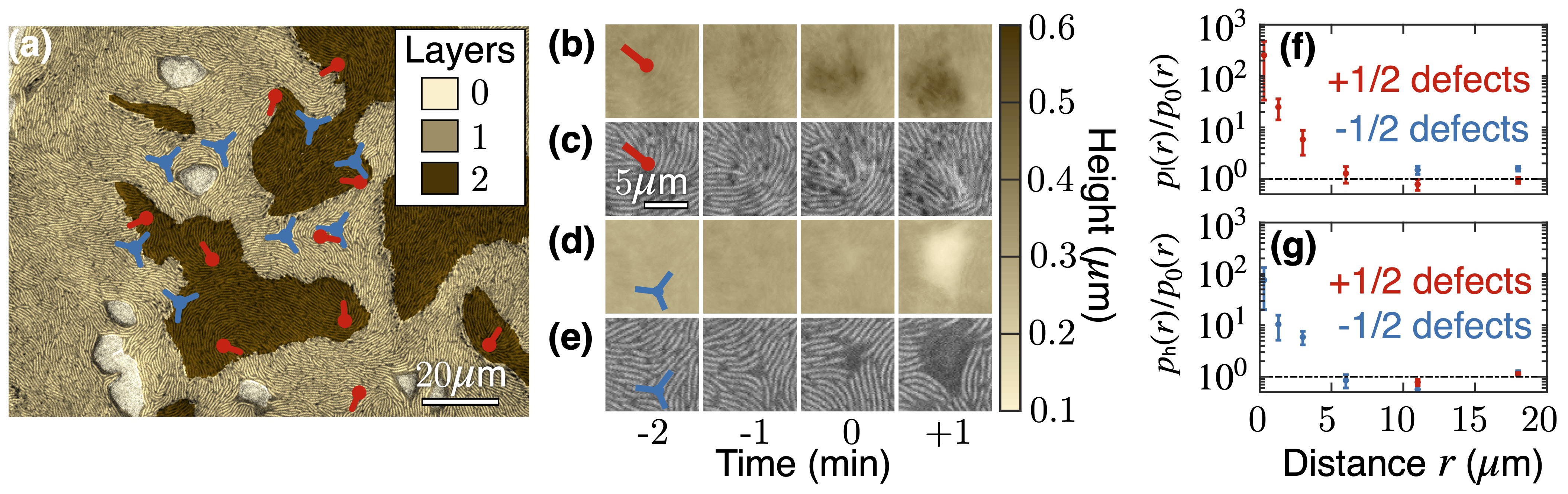}
  {\phantomsubcaption\label{defect-tracking}}
  {\phantomsubcaption\label{height+1/2}}
  {\phantomsubcaption\label{layer+1/2}}
  {\phantomsubcaption\label{height-1/2}}
  {\phantomsubcaption\label{layer-1/2}}
  {\phantomsubcaption\label{probability-layers}}
  {\phantomsubcaption\label{probability-holes}}
    \bfcaption{New layers and new holes preferentially form at $+1/2$ and $-1/2$ topological defects, respectively}{ (\subref*{defect-tracking}) Layer-count field overlaid on a reflectance image of the bacterial colony. $+1/2$ and $-1/2$ defects are marked with red and blue symbols, respectively. (\subref*{height+1/2}-\subref*{layer-1/2}) Snapshots of the height and the reflectance fields for both a layer-formation event at a $+1/2$ defect (\subref*{height+1/2}, \subref*{layer+1/2}), and a hole-opening event at a $-1/2$ defect (\subref*{height-1/2}, \subref*{layer-1/2}). (\subref*{probability-layers},\subref*{probability-holes}) Probability distributions of the distance between defects and new layers (\subref*{probability-layers}), and between defects and new holes (\subref*{probability-holes}). Following Ref. \cite{Saw2017}, we normalized the distributions by the distribution $p_0(r)$ of distances between defects and randomly selected points within the monolayer (excluding holes, see alternative normalization in \cref{alternative-normalization}). Therefore, the plots indicate how much more likely it is for new layers and holes to appear at a given distance from $+1/2$ and $-1/2$ defects. Histograms are built based on 134 $+1/2$-to-layer and 121 $-1/2$-to-layer distances from 35 layer formation events, and on 402 $+1/2$-to-hole and 318 $-1/2$-to-hole distances from 59 hole opening events. Errors are S.D.}
\end{figure*}

Densely-packed, motile, rod-shaped objects can form a state of matter called an active nematic liquid crystal \cite{Doostmohammadi2018}. Active nematics display emergent collective phenomena resulting from the interplay between active stresses, here due to cell motility, and orientational order, here due to mechanical cell-cell interactions \cite{Doostmohammadi2018,Marchetti2013,Aranson2019,Bar2020,Sengupta2020}. To quantify orientational order in the \textit{M. xanthus} colony, we used the reflectance field to measure both the degree of alignment $S(\bm{r},t)$ and the cell orientation angle $\theta(\bm{r},t)$ (\cref{angle-field,strength-field}, \hyperref[methods]{Methods}, \hyperref[movies]{Movies S4 and S5}). The orientation angle varies smoothly throughout most of the colony, with a correlation length $\ell_{\text{n}} \approx 16.0\pm0.5$ $\mu$m (\cref{correlations}). However, at a discrete set of points known as topological defects all orientations meet and hence the orientation angle is singular (\cref{angle-field,angle-closeup}). Alignment is lost at the defect points, which are furthermore often connected by lines with lower alignment than the perfectly-ordered background (\cref{strength-field}). Consistent with the nematic symmetry of cell alignment, we observe point defects with half-integer topological charges of $\pm 1/2$ (\cref{angle-closeup,angle+1/2,angle-1/2}) \cite{deGennes-Prost}. Whereas $+1/2$ defects have one axis of symmetry (red segment, \cref{angle+1/2}), $-1/2$ defects have three axes of symmetry (blue segments, \cref{angle-1/2}). As expected in an active nematic \cite{Doostmohammadi2018,Giomi2013,Shi2013}, defects are spontaneously created and annihilated either in oppositely-charged pairs or individually at boundaries of a layer or hole (\hyperref[movies]{Movie S6}).

Similar defect dynamics have been found in other systems including vibrated granular rods \cite{Narayan2007}, mixtures of cytoskeletal filaments and molecular motors \cite{Sanchez2012,Kumar2018}, monolayers of mesenchymal and epithelial cells \cite{Duclos2017,Kawaguchi2017,Saw2017,Blanch-Mercader2018}, growing bacterial colonies \cite{Doostmohammadi2016a,DellArciprete2018,Yaman2019,Echten2020}, and colonies of swarming filamentous bacteria \cite{Li2019}. In some cell populations, topological defects influence collective cell motion and can even trigger intracellular responses \cite{Saw2018}. Both in suspensions of bacteria swimming in passive liquid crystals \cite{Peng2016,Genkin2017}, and in mesenchymal cell monolayers \cite{Kawaguchi2017,Endresen2019,Turiv2020}, cells were found to accumulate around positive defects and become depleted from negative defects. In chaining bacterial biofilms, stress accumulation at $-1/2$ defects was found to induce a buckling instability that leads to sporulation \cite{Yaman2019}. Further, in epithelial monolayers, increased pressure around $+1/2$ defects was found to induce cell apoptosis and extrusion \cite{Saw2017}. Respectively, in mesenchymal monolayers, compressive stress around integer defects triggers cell differentiation \cite{Guillamat2020}. Finally, topological defects in the nematic order of supracellular actin fibers have been recently found to organize \textit{Hydra} morphogenesis \cite{Maroudas-Sacks2020}.

Here, we find that topological defects play an important role in the developmental cycle of \textit{M. xanthus}: they promote the layering process that leads to fruiting body formation. We observed new layers forming stochastically throughout the colony. However, by identifying and tracking topological defects (\cref{defect-tracking}, \hyperref[movies]{Movie S6}, \hyperref[methods]{Methods}), we found many events in which new cell layers form close to $+1/2$ defects (\cref{height+1/2,layer+1/2}), and new holes open close to $-1/2$ defects (\cref{height-1/2,layer-1/2}). To quantify this relationship, we measured the distribution of distances between defects of either sign and the locations where new layers and new holes appear. These measurements indicate that it is $\sim 200$ times more likely for a new layer to form close to a $+1/2$ defect than away from it (\cref{probability-layers}), and that it is $\sim 80$ times more likely for a new hole to open at a $-1/2$ defect than away from it (\cref{probability-holes}). Unlike in growing biofilms, where pressure induces layering and verticalization transitions at a critical colony size \cite{Su2012,Grant2014,Duvernoy2018,Beroz2018,Warren2019,Yaman2019,You2019}, we observe local migration-induced layering events independent of colony size.

\begin{figure*}[t]
    \includegraphics[width=\textwidth]{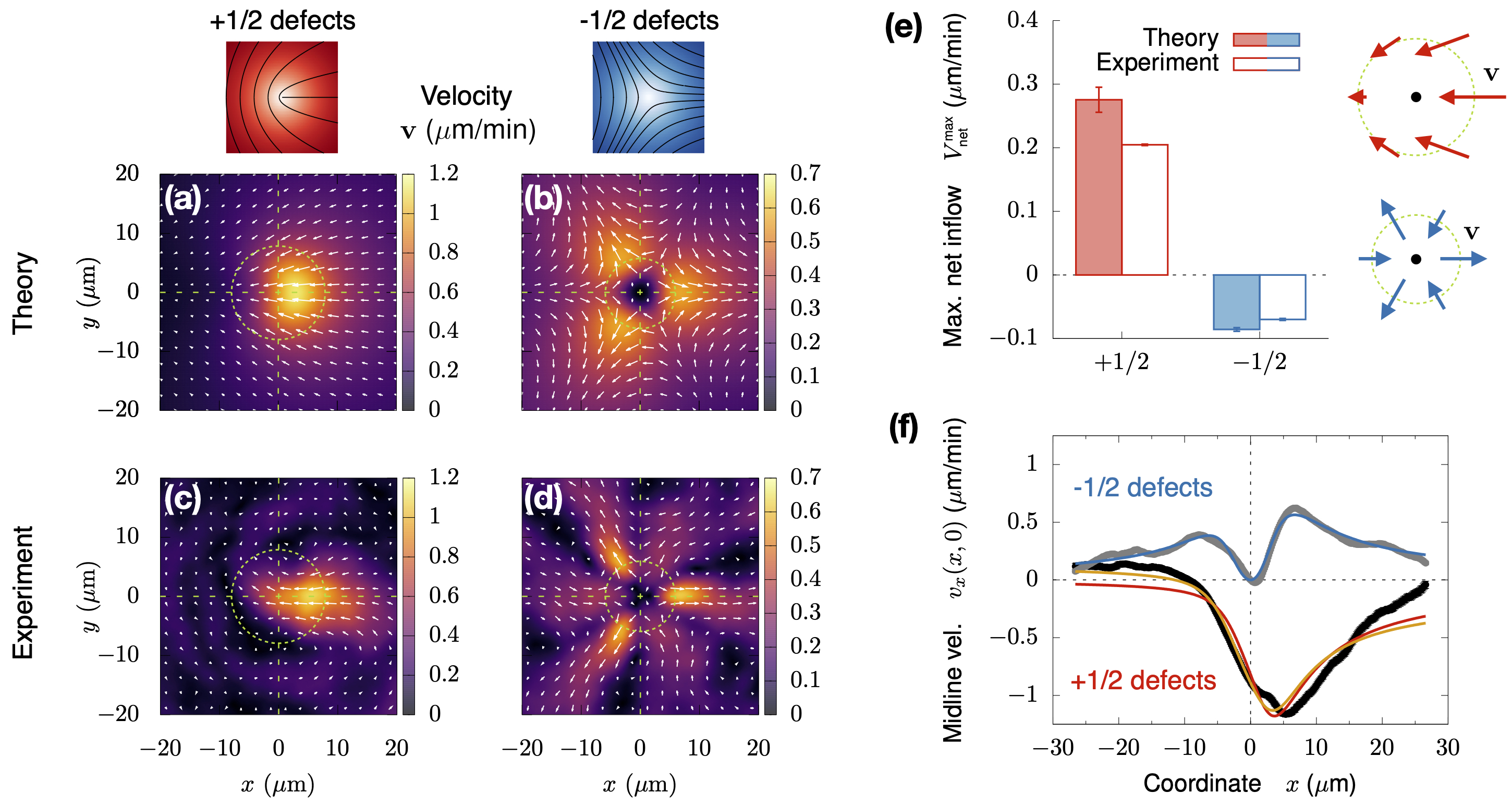}
  {\phantomsubcaption\label{+1/2-flow-theory}}
  {\phantomsubcaption\label{-1/2-flow-theory}}
  {\phantomsubcaption\label{+1/2-flow-experiment}}
  {\phantomsubcaption\label{-1/2-flow-experiment}}
  {\phantomsubcaption\label{net-inflow}}
  {\phantomsubcaption\label{midline-flow-fits}}
    \bfcaption{Asymmetric cell flows around topological defects explain the formation of new layers and holes}{ The defect schematics show the order parameter (color map) along with director-field lines. Theoretically predicted (\subref*{+1/2-flow-theory}, \subref*{-1/2-flow-theory}) and experimentally measured (\subref*{+1/2-flow-experiment}, \subref*{-1/2-flow-experiment}) flow fields around topological defects. The color maps indicate the speed, and the white arrows indicate the velocity. The defect cores are located at the origin of coordinates. Averages are over a total of 7896 frames from 96 tracks of $+1/2$ defects and 7096 frames from 144 tracks of $-1/2$ defects, across 8 replicate experiments (\hyperref[methods]{Methods}). See the average flow fields measured in individual replicate experiments in \cref{replicates}. (\subref*{net-inflow}) Maximal net inflow of cells at $+1/2$ defects ($V^{\text{max}}_{\text{net}}>0$, red), and net outflow from $-1/2$ defects ($V^{\text{max}}_{\text{net}}<0$, blue), as seen in the coarsely-sampled predicted flow fields around defects shown on the right. $V_{\text{net}}(R)$, defined in \cref{eq net-inflow}, is the average net inflow velocity through a circumference of radius $R$ centered at a defect (\cref{net-inflow-radius}). $V^{\text{max}}_{\text{net}}$ is the maximum of $V_{\text{net}}(R)$ in the experiments, which occurs at $R=8$ $\mu$m for $+1/2$ defects and at $R=6$ $\mu$m for $-1/2$ defects (dashed green circumferences in panels \subref*{+1/2-flow-theory} to \subref*{-1/2-flow-experiment}). The theoretical $V_{\text{net}}(R)$ is given in \cref{eq net-inflow-prediction}, with error bars obtained by propagation of errors in the parameter values (\cref{t fits}). Errors are S.E.M. (\subref*{midline-flow-fits}) Fits of the theoretical predictions to the experimentally measured velocity profiles along the midline ($y=0$) of $+1/2$ (black data points and red fitted curve) and $-1/2$ (grey data points and blue fitted curve) defects. The optimal values of the fitting parameters, which were used to plot panels \subref*{+1/2-flow-theory},\subref*{-1/2-flow-theory}, and \subref*{net-inflow}, are listed in \cref{t fits}. The yellow curve is the fit of the extended model that accounts for the counterflow measured in front of $+1/2$ defects, in the region $x\lesssim -10$ $\mu$m. The corresponding optimal parameter values are listed in \cref{t fits-compressible}. Error bars (S.E.M.) are barely visible because they are comparable to the point size.}
\end{figure*}

To understand the association between topological defects and layering, we modeled the cell colony as a thin film of active nematic fluid (\hyperref[theory]{Supplementary Text}). We describe cell alignment in terms of the nematic order parameter tensor field $\bm{Q}(\bm{r})$, which we assume relaxes rapidly to its equilibrium configuration. As cells migrate along the alignment axis, they mechanically interact with neighboring cells, which produces an anisotropic active stress $\bm{\sigma}^{\text{a}} = -\zeta \bm{Q}$ in the colony. The coefficient $\zeta$ is positive (negative) for extensile (contractile) stresses. The distortions of cell alignment found around topological defects give rise to a non-zero active force density $\bm{f}^{\text{a}} = \bm{\nabla}\cdot \bm{\sigma}^{\text{a}}$, which drives cell flows $\bm{v}(\bm{r})$ \footnote{Because the height of the fluid layer can change freely, the two-dimensional velocity field $\bm{v}(\bm{r})$ can have a non-zero divergence, $\bm{\nabla}\cdot\bm{v} \neq 0$ (see \cref{eq height} in the \hyperref[theory]{Supplementary Text}).}. The active forces are balanced by viscous friction forces $\bm{\xi}\bm{v}$ arising from cell-substrate interactions:
\begin{equation} \label{eq force-balance-main}
\bm{\xi}\bm{v} = \bm{f}^{\text{a}}.
\end{equation}
Like the active stresses, we assume that friction forces are anisotropic. Following Ref. \cite{Kawaguchi2017}, we account for friction anisotropy via a friction coefficient matrix $\bm{\xi} = \xi_0 \left[ \mathbb{I} - \epsilon\, \bm{Q}\right]$, where the first term corresponds to isotropic friction with coefficient $\xi_0$, and $\epsilon$ is the friction anisotropy along the local alignment axis. Previous measurements of the mechanical response of cell-substrate focal adhesions in \textit{M. xanthus} \cite{Balagam2014} suggest that friction is smaller along the cell-alignment axis than perpendicular to it, i.e. $\epsilon>0$.

Assuming equilibrium solutions for the order parameter $\bm{Q}(\bm{r})$ around topological defects, we solved \cref{eq force-balance-main} to predict the cell flow fields around $+1/2$ and $-1/2$ defects (\hyperref[theory]{Supplementary Text}, \cref{eq flow-field+1/2,eq flow-field-1/2}). The results show that $+1/2$ defects exhibit net self-driven motion along their axis of symmetry (\cref{+1/2-flow-theory}), a well-known feature of active nematic fluids \cite{Doostmohammadi2018,Giomi2013,Giomi2014a,Shi2013,Pismen2013,Shankar2018b}. Moreover, cells in front of the defect are aligned perpendicularly to the flow, thus experiencing stronger friction than cells behind the defect. Therefore, due to the positive friction anisotropy $\epsilon>0$, the inflow toward the defect core is stronger than the outflow (\cref{+1/2-flow-theory}). As a result, cells accumulate at $+1/2$ defects, and are eventually extruded vertically to form new cell layers (\cref{height+1/2,layer+1/2}). For $-1/2$ defects, anisotropic friction gives rise to a stronger outflow than inflow (\cref{-1/2-flow-theory}), which explains the opening of holes at $-1/2$ defects (\cref{height-1/2,layer-1/2}). Were friction isotropic, the velocity field around defects would be symmetric, implying no cell accumulation or depletion, and hence no preferential layer formation or hole opening at defects.

To test our predictions, we measured cell flows around topological defects by calculating optical flow from the reflectance movies (\hyperref[methods]{Methods}). In agreement with our theoretical predictions, the measured flow fields show that $+1/2$ defects self-propel along their axis, and that there is a net inflow toward $+1/2$ defects, and a net outflow from $-1/2$ defects (\cref{+1/2-flow-experiment,-1/2-flow-experiment,net-inflow,net-inflow-radius}). Also in agreement with our predictions, we find that cell accumulation ($\bm{\nabla}\cdot \bm{v}<0$) occurs mainly in front of $+1/2$ defects, whereas cell depletion ($\bm{\nabla}\cdot \bm{v}>0$) is localized in three lobes along the axes of symmetry of $-1/2$ defects (\cref{divergence}).

To quantitatively assess our model, we fit the theoretical predictions to the measured velocity profile along the midline of topological defects, i.e. $v_x(x,y=0)$ (\cref{midline-flow-fits}, \hyperref[theory]{Supplementary Text}). The fits yield values for three model parameters: the nematic healing length $\ell$ that determines the defect core size, the ratio $\zeta/\xi_0$ between active stress and isotropic friction coefficients, and the friction anisotropy $\epsilon$ (\cref{t fits}). For both $+1/2$ and $-1/2$ defects, we obtain $\ell\approx 2.5$ $\mu$m; the defect core is smaller than a cell length $l \approx 7$ $\mu$m. In addition, we find that active stress in the bacterial colony is extensile ($\zeta>0$), and that friction anisotropy is positive ($\epsilon>0$). However, the optimal values of $\zeta/\xi_0$ and $\epsilon$ are different for $+1/2$ and $-1/2$ defects. These effective parameter values characterize the average flows around defects, not all of which give rise to new layers or holes. Imposing common fitting parameters for $+1/2$ and $-1/2$ defects, we obtain poorer fits (\cref{simultaneous-fits,t simultaneous-fits}). To obtain compatible parameter values for both defect types, future detailed models of layer formation and hole opening might have to account for additional forces such as the surface tension of the water layer that wets the colony.

For $-1/2$ defects, the fit (blue curve) closely matches the data (grey points, \cref{midline-flow-fits}). However, in front of $+1/2$ defects, the fit (red curve) qualitatively departs from the data (black points, \cref{midline-flow-fits}). Whereas for $+1/2$ defects our model predicts a negative velocity $v_x<0$ everywhere along the defect midline, we measure a positive velocity $v_x>0$ for $x\lesssim -10$ $\mu$m (\cref{midline-flow-fits}), indicating that cells in this region flow toward the defect core, opposite to the overall motion of the defect. In the following, we propose a possible explanation for this observation. Our model so far assumed that, under compression, cells are readily extruded from the cell monolayer. Accordingly, we assumed that the cell density is uniform ($\rho(\bm{r})=\rho_0$) and, hence, we neglected pressure gradients in \cref{eq force-balance-main}. However, if cells are not readily extruded, the non-uniform flows around a $+1/2$ defect produce compression in the cell monolayer, leading to a pressure increase in front of $+1/2$ defects. The resulting pressure gradient drives an additional flow toward the defect core, which might account for the measured net counterflow. To probe this idea, we generalized our active nematic model to allow for a non-uniform cell density $\rho(\bm{r})$ and the associated pressure gradients (\hyperref[theory]{Supplementary Text}). The force balance then reads
\begin{equation}
\bm{\xi}\bm{v} = -\bm{\nabla} P (\rho) + \bm{f}^{\text{a}},
\end{equation}
and we assume a linear equation of state $P(\rho) = B\left[\rho-\rho_0\right]/\rho_0 + P(\rho_0)$, where $B$ is the bulk modulus of the monolayer. With additional assumptions about the rate of cell extrusion (\hyperref[theory]{Supplementary Text}), this extended model produces counterflow, as shown by the fit of the predicted midline velocity (yellow curve in \cref{midline-flow-fits}, \hyperref[theory]{Supplementary Text}). Gaining further insight into the physical origin of the counterflow will require future experiments that can accurately measure the cell density and mechanical stress around defects, for example employing cell segmentation algorithms and traction force microscopy, respectively.

In summary, a dense monolayer of \textit{M. xanthus} cells migrating on a surface forms an active nematic liquid crystal. Spontaneously-created topological defects of the liquid-crystalline alignment promote the formation of new cell layers and holes in the bacterial colony. The localization of these layering events at topological defects results from the anisotropic friction that the cells experience as they migrate. Thus, we propose that cell motility and mechanical interactions between cells are sufficient to drive the formation of multilayered structures. We have performed our experiments in the presence of nutrients. However, the layering mechanism that we have uncovered is a generic feature of dense active nematics with anisotropic friction on a substrate. Therefore, we expect that the same mechanism is responsible for layer formation when nutrients are scarce. Under starvation, \textit{M. xanthus} cells alter their motility to drive aggregation of the colony into dense cell monolayers, from which fruiting bodies form \cite{Liu2019a}. Future work will be needed to connect starvation-induced changes in motility to the full development of stable fruiting bodies.

Finally, the formation of multilayered cell structures from an initial cell monolayer is not unique to \textit{M. xanthus}: it occurs both in biofilms of other bacterial species \cite{Su2012,Grant2014,Duvernoy2018,Shrivastava2018,Warren2019,You2019,Takatori2020}, as well as in the stratification of the mammalian epidermis \cite{Miroshnikova2018}. Even though the precise cellular mechanisms may differ in different systems, it is appealing to speculate that population-scale orientational order and the associated topological defects may provide a generic route to the formation of cell layers in a wide range of biological systems.



\section*{Acknowledgments}
We thank Farzan Beroz, Matthew Black, Chenyi Fei, Endao Han, M. Cristina Marchetti, John McEnany, and Cassidy Yang for discussions. This work was supported in part by the National Science Foundation, through award PHY-1806501 (J.W.S.) and the Center for the Physics of Biological Function (PHY-1734030), and by the National Institutes of Health award R01GM082938 (N.S.W.). R.A. acknowledges support from the Human Frontiers of Science Program (LT000475/2018-C). The authors acknowledge the use of Princeton's Imaging and Analysis Center, which is partially supported by the Princeton Center for Complex Materials, a National Science Foundation (NSF)-MRSEC program (DMR-1420541).

\section*{Author contributions}
K.C. performed the experiments and analyzed the data. R.A. developed the theory and fitted the predictions to the experimental data. All authors interpreted the results and designed experiments. N.S.W. and J.W.S. supervised the study. K.C. and R.A. wrote the manuscript with input from all authors.

\section*{Competing interests}
The authors declare no competing interests.

\section*{Data availability}
All data are available from the authors upon request.

\section*{Code availability}
All codes are available from the authors upon request.

\bibliography{Bacterial_layers}

\onecolumngrid 

\clearpage

\onecolumngrid
\begin{center}
\textbf{\large Supporting Information for}\\
\smallskip
\textbf{\large ``Topological defects promote layer formation in \textit{Myxococcus xanthus} colonies''}
\bigskip
\end{center}

\setcounter{equation}{0}
\setcounter{figure}{0}
\renewcommand{\theequation}{S\arabic{equation}}
\renewcommand{\thefigure}{S\arabic{figure}}

\twocolumngrid

\section*{Materials and Methods} \label{methods}

\subsection*{Cell culture}

\textit{Myxococcus xanthus} cells of the wild-type strain DK1622 are grown in CTTYE ($1\%$ Casitone, $10$ mM Tris-HCl [$\mathrm{pH} = 7.6$], $1$ mM KH$_2$PO$_4$, $8$ mM MgSO$_4$) overnight to a concentration corresponding to a range of OD$_{600} \approx 0.4 - 0.8$, where OD$_{600}$ is the optical density at a wavelength of $600$ nm. This culture is then concentrated and re-suspended into fresh CTTYE to an OD$_{600}\approx 2$. Then, $5$ $\mu$l of concentrated culture is placed onto a CTTYE pad with $1.5\%$ agarose, and allowed to dry until no liquid is visible on the surface of the agarose gel. Next, the sample is incubated for 2 hours before imaging.

\subsection*{Imaging}

We image our samples using a Keyence VK-X 1000 microscope. The microscope works by scanning a laser across the surface of the sample and measuring the reflectance of the laser light on the surface. The reflectance of the sample is measured at multiple heights. Then, a height field of the sample (\cref{height}) is obtained by finding, for each pixel, the height at which the reflectance is maximal. Finally, the reflectance field at the height of maximal reflectance gives clear and in-focus images of the top of sample (\cref{brightness}). Imaging was done at a frame rate between $4.8$ and $7.2$ min$^{-1}$ for 1 to 3 hours. During this time, the colony does not grow substantially. At the temperature of the experiments, $T=22\degree$C, the cell doubling time is $\approx 10$ h \cite{Janssen1977}. The entire cell colony is over $1$ cm in diameter; our field of view is a region of $\sim 150\times 100$ $\mu$m. Data in this manuscript come from 8 replicate experiments.

\subsection*{Cell motility measurements and state of the colony}

\textit{M. xanthus} colonies undergo a transition from a non-aggregated state in which the colony forms a thin film on the agar substrate to an aggregated state that leads to fruiting body formation. A recent study has shown that this transition is controlled by a dimensionless number known as the inverse rotational P\'{e}clet number,
\begin{equation}
\mathrm{Pe}_{\text{r}}^{-1} = \frac{\ell_{\text{cell}}}{v_0}(D_{\text{r}} + 2 f_{\text{rev}}),
\end{equation}
which combines the effects of the cells' self-propulsion speed $v_0$, their rotational diffusion coefficient $D_{\text{r}}$, and the average rate of velocity reversals $f_{\text{rev}}$ \cite{Liu2019a}. $\ell_{\text{cell}} = 2.5$ $\mu$m is an effective average cell size \cite{Liu2019a}.

Here, in order to study layer formation, we perform experiments in the presence of nutrients, which maintains the cell colony in the regime of cell migration parameters corresponding to the non-aggregated state. In this state, fruiting bodies do not form, but new cell layers and holes continuously appear and disappear, providing us with many layer formation and hole opening events that allow us to study these processes in detail. If we performed the experiments in starvation conditions, the colony would transition to the aggregated state and fruiting bodies would rapidly form, preventing us from imaging the layer formation process in sufficient detail.

\begin{figure}
    \includegraphics[width=\columnwidth]{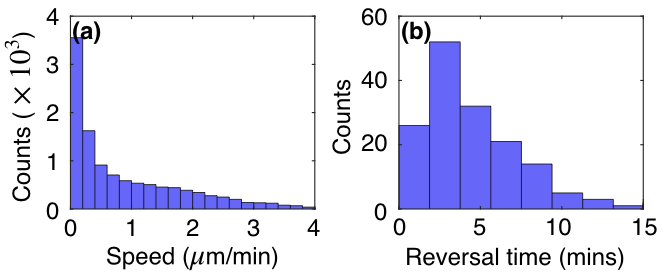}
  {\phantomsubcaption\label{speed}}
  {\phantomsubcaption\label{reversals}}
    \bfcaption{Statistics of \textit{M. xanthus} motility}{ Histograms of the cell speed (\subref*{speed}) and time between velocity reversals (\subref*{reversals}) of cells migrating in a low-cell-density environment.}
    \label{motility}
\end{figure}

To check that our experiments lie within the appropriate parameter regime, we tracked cells moving at a low cell density corresponding to an OD$_{600} = 0.1$, and we measured the probability distribution of cell speeds and of times between velocity-reversal events (\cref{motility}). We find that cells move with an average speed $v_0 = 0.922 \pm 0.009$ $\mu$m/min, and an average reversal rate $f_{\text{rev}} = 0.172 \pm 0.002$ min$^{-1}$. The value of the rotational diffusion coefficient, $D_r = 0.127$ min$^{-1}$, was taken from Ref. \cite{Liu2019a}. These motility parameter values give $\mathrm{Pe}_{\text{r}}^{-1} = 1.28 \pm 0.09$, which lies well within the non-aggregated state in the phase diagram of Ref. \cite{Liu2019a}.

\subsection*{Nematic order}

To characterize the nematic order in the cell colony, we measure the orientation angle field $\theta(\bm{r},t)$ and the nematic order parameter strength field $S(\bm{r},t)$ from the laser brightness images of the colony. Following Ref. \cite{Li2019}, we first sharpen the laser-brightness signal by means of a high-pass filter. Then, we compute the gradient of the resulting brightness field $I_{ij}$, where the indices indicate the image pixels. We smooth the brightness gradient by means of a Gaussian filter with standard deviation $\sigma = 1.7$ $\mu$m. Then, for each pixel, we build the so-called structure tensor
\begin{equation}
H_{ij} = 
\begin{pmatrix}
(\partial_x I_{ij})^2 & (\partial_x I_{ij})(\partial_y I_{ij}) \\
(\partial_x I_{ij})(\partial_y I_{ij}) & (\partial_y I_{ij})^2
\end{pmatrix},
\end{equation}
and we obtain its eigenvalues and eigenvectors. The eigenvector associated with the smallest eigenvalue gives the local direction of least brightness gradient, which we identify with the angle field $\theta_{ij}$ of cell orientation.

From the measured angle field $\theta_{ij}$, we obtain the order parameter strength field $S_{ij}$ as \cite{deGennes-Prost}
\begin{equation}
S_{ij} = \langle 2\cos(\theta_{ij}-\bar\theta_{ij}) - 1 \rangle_R,
\end{equation}
where $\bar\theta_{ij} = \langle \theta_{ij}\rangle_R$ is the average orientation angle within a disk of radius $R=1$ $\mu$m centered at pixel $(i,j)$. $S_{ij}$ is almost $1$ over the majority of the system, indicating an almost perfect cell alignment, but it approaches $0$ at defects, where cell alignment is lost and the angle field is singular (\cref{strength-field}).

\begin{figure}
    \includegraphics[width=\columnwidth]{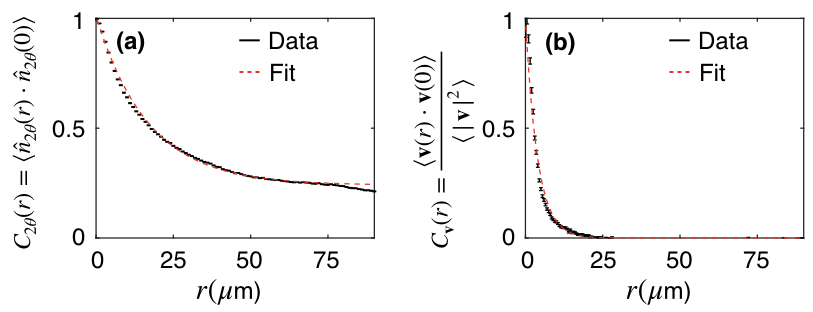}
  {\phantomsubcaption\label{nematic-correlation}}
  {\phantomsubcaption\label{velocity-correlation}}
    \bfcaption{Correlation functions}{ Spatial correlation functions of (\subref*{nematic-correlation}) the modified director field $\hat{\bm{n}}_{2\theta} = (\cos(2\theta),\sin(2\theta))$, and (\subref*{velocity-correlation}) of the velocity field $\bm{v}$. Error bars are S.E.M. By fitting exponential decays, we obtain the nematic and the velocity correlation lengths: $\ell_{\text{n}} = 16.0\pm0.5$ $\mu$m, and $\ell_{\text{v}} = 3.9\pm0.1$ $\mu$m, respectively.}
    \label{correlations}
\end{figure}

Following Ref. \cite{Li2019}, we measure the following spatial correlation function of the nematic orientation angle,
\begin{equation} \label{eq nematic-correlation-function}
C_{2\theta}(r) = \langle \hat{\bm{n}}_{2\theta} (r)\cdot \hat{\bm{n}}_{2\theta}(0)\rangle,
\end{equation}
where $\hat{\bm{n}}_{2\theta} = (\cos (2\theta), \sin(2\theta))$. The average runs over 10000 randomly-selected pairs of points from every frame of every experiment. We fit an exponential decay with an offset $C_{2\theta}(r) = A\, e^{-r/\ell_{\text{n}}} + B$ to the measured correlation function (\cref{nematic-correlation}) and obtain the nematic correlation length $\ell_{\text{n}} = 16.0\pm0.5$ $\mu$m.

Finally, to measure the equilibrium value of the order parameter strength, $S_0$ (see \hyperref[theory]{Supplementary Text}), we obtain the average $S_0 = \langle S(\bm{r},t)\rangle$ over regions of the cell colony without defects. Specifically, we exclude circular regions of radius $R_{\text{def}} = 16$ $\mu$m $\approx \ell_{\text{n}}$ around topological defects (see below for a description of the defect detection method). In addition, we also exclude holes in the cell colony, i.e. regions without cells. Using this procedure, we obtain $S_0 = 0.992\pm 0.003$.

\subsection*{Defect detection and tracking}

We identify and track defects by finding local minima of the nematic order parameter strength $S(\bm{r},t)$ with $S<0.7$. Often, several pixels around these local minima have the same value of $S$. Thus, we identify the defect core with the centroid of all the pixels that share the local minimum value of $S$.

For each point identified as a potential defect core, we calculate the topological charge $q$ by applying its definition:
\begin{equation}
q = \frac{1}{2\pi} \oint_{\mathcal{C}} \dd \theta,
\end{equation}
where $\mathcal{C}$ is a circular circuit of radius $R_{\text{c}} = 1$ $\mu$m around the defect core. We ignore the candidate points with $q=0$, as well as the ones that fall in holes in the cell colony.

For each defect, we identify its axes of symmetry as recently proposed in Ref. \cite{Vromans2016}. Specifically, for defects with topological charge $+1/2$, the defect axis is given by $\hat{\bm{p}} = \bm{\nabla}\cdot\bm{Q}/|\bm{\nabla}\cdot\bm{Q}|$, where $\bm{Q}$ is the nematic order parameter tensor with components $Q_{\alpha\beta} = S [ 2\, \hat{n}_\alpha \hat{n}_\beta - \delta_{\alpha\beta}]$, with $\hat{\bm{n}} = (\cos\theta,\sin\theta)$ the director field. Respectively, for defects with topological charge $-1/2$, we first define the opposite nematic angle $\theta' = -\theta$, and we obtain the corresponding $\bm{Q}'$ tensor. From it, we obtain $\hat{\bm{p}}' = \bm{\nabla}\cdot\bm{Q}'/|\bm{\nabla}\cdot\bm{Q}'|$. Finally, defining $\hat{\bm{p}}' = (\sin \psi',\cos\psi')$, one symmetry axis of a $-1/2$ defect is given by the angle $\psi = -\psi'/3$. The other two symmetry axes of a $-1/2$ defect are then defined by the three-fold symmetry of $-1/2$ defects.

Finally, we track defects over time by finding the closest defects with the same charge in consecutive image frames. We ignore defects that last less than $10$ frames ($\approx 2$ min). Across 8 replicate experiments, we found $96$ defects with a topological charge of $q = +1/2$ and $144$ defects with $q = -1/2$. All together, the $+1/2$ defect tracks consist of a total of $7896$ defect frames, which are included in the averaged data in \cref{angle+1/2,+1/2-flow-experiment}. Respectively, a total of $7096$ $-1/2$ defect frames are included in the averaged data in \cref{angle-1/2,-1/2-flow-experiment}.

\begin{figure}
    \includegraphics[width=0.5\columnwidth]{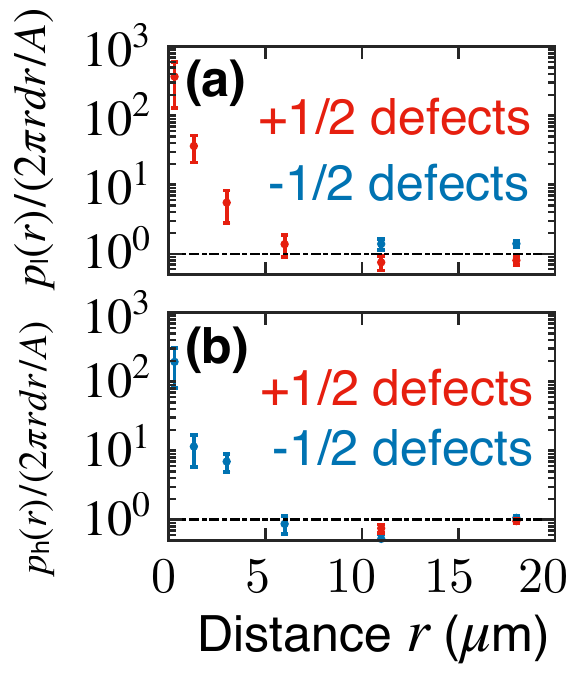}
  {\phantomsubcaption\label{defects-layers}}
  {\phantomsubcaption\label{defects-holes}}
    \bfcaption{Probability distributions of the distance between defects and new layers (\subref*{defects-layers}), and between defects and new holes (\subref*{defects-holes})}{ In \cref{probability-layers,probability-holes}, we normalized these distributions with the distribution $p_0(r)$ of distances between defects and randomly selected points within the monolayer (excluding holes). Here, as commonly done for radial distribution functions $g(r)$, we normalized the distributions by the area of an annulus of width $dr$, the histogram bin size, with $A$ the area of the field of view. Both normalizations give similar results. Errors are S.D.}
    \label{alternative-normalization}
\end{figure}

\begin{figure*}
    \includegraphics[width=0.7\textwidth]{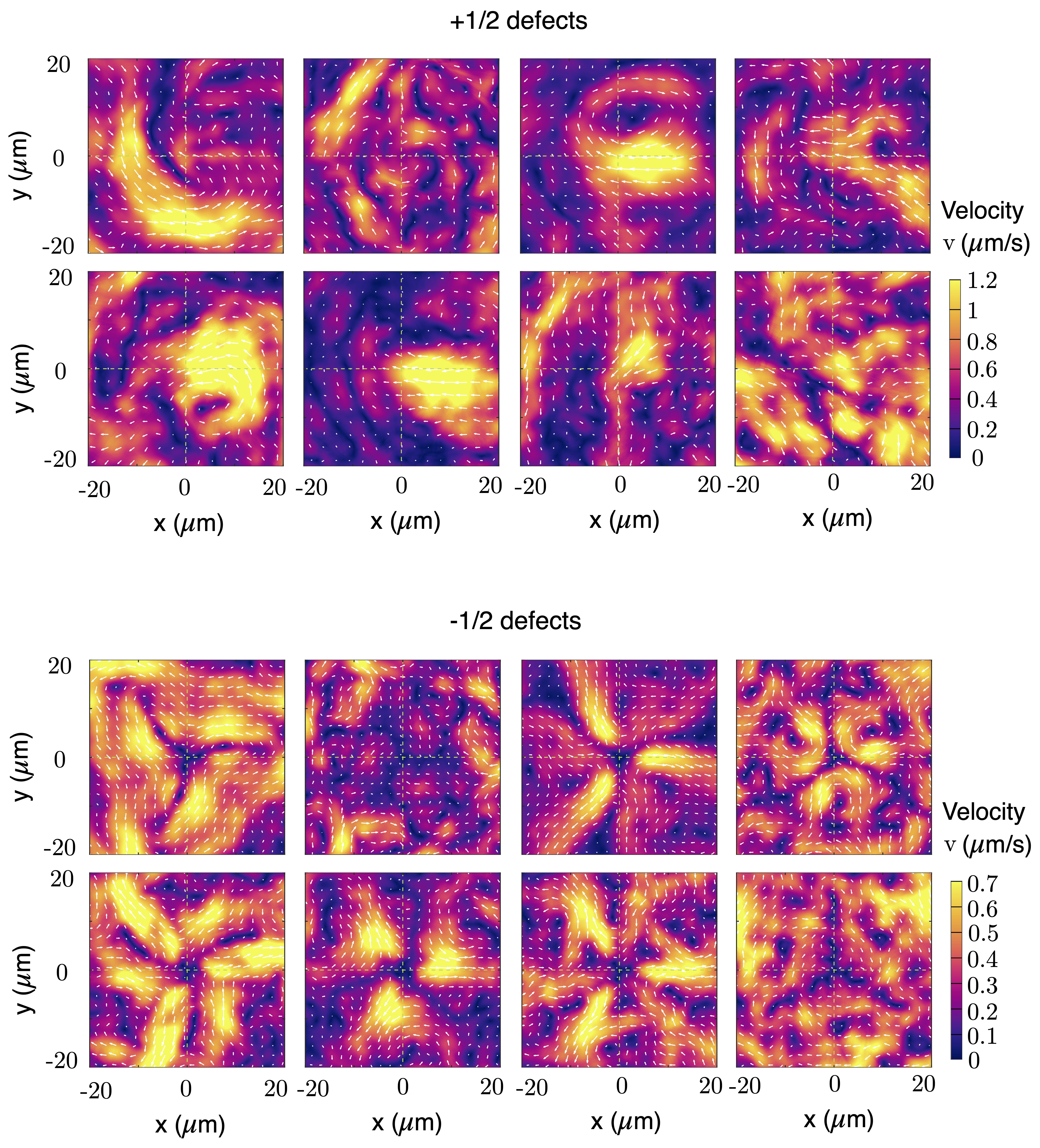}
    \bfcaption{Average flow field around defects separately measured in each of the 8 replicate experiments}{ Each of the separate averages includes a different number of defect frames. The total averages in \cref{+1/2-flow-experiment,-1/2-flow-experiment} include all the defect frames across the 8 replicate experiments.}
    \label{replicates}
\end{figure*}

\subsection*{Cell flow measurements}

We measure cell flows by applying an optical-flow algorithm on the laser-brightness images of the cell colony. The optical-flow algorithm is based on the Lucas-Kanade derivative of Gaussian method \cite{Lucas1981}. By comparing independent velocity measurements from a small number of manual cell trackings to the optical-flow measurements, we find a calibration parameter to convert the optical flow to a cell velocity field within the bacterial colony.

Finally, we measure the spatial correlation function of the velocity field:
\begin{equation}
C_{\text{v}}(r) = \frac{\langle\bm{v}(r)\cdot\bm{v}(0)\rangle}{\langle |\bm{v}|^2\rangle}.
\end{equation}
As for the nematic correlation function \cref{eq nematic-correlation-function}, the average runs over 10000 randomly-selected pairs of points from every frame of every experiment. We fit an exponential decay $C_{\text{v}}(r) = D\, e^{-r/\ell_{\text{v}}}$ to the measured correlation function (\cref{velocity-correlation}) and obtain the velocity correlation length $\ell_{\text{v}} = 3.9\pm0.1$ $\mu$m.

\section*{Supplementary Movies} \label{movies}

\noindent\textbf{Movie S1.} Reflectance field of a thin cell colony, with new holes and cell layers appearing and disappearing.

\smallskip

\noindent\textbf{Movie S2.} Height field corresponding to Movie S1.

\smallskip

\noindent\textbf{Movie S3.} Reflectance field of a thin colony of $\Delta$\textit{pilA} mutant cells that lack pili. As for the wild-type cells, colonies of pili-lacking cells also produce new holes and new cell layers.

\smallskip

\noindent\textbf{Movie S4.} Color map of the cell orientation angle $\theta$ overlaid on a magnified reflectance movie of the colony.

\smallskip

\noindent\textbf{Movie S5.} Nematic alignment field $S$ of the colony.

\smallskip

\noindent\textbf{Movie S6.} Red and blue symbols track the positions and orientations of $+1/2$ and $-1/2$ topological defects, respectively, as they spontaneously appear, move, and annihilate within the cell colony. The color map shows the number of layers overlaid on a laser-brightness movie of the colony. The color code is as in \cref{layers}.

\onecolumngrid

\clearpage

\twocolumngrid

\section*{Supplementary Text} \label{theory}


\subsection{Active nematic model of the bacterial colony} \label{model}

In this section, we model a monolayer of \textit{Myxococcus xanthus} cells as an active nematic fluid. The cell monolayer forms a phase with nematic liquid-crystalline order as a result of the dense packing and rod-like shape of \textit{M. xanthus} cells. The nematic nature of the ordering is confirmed by the presence of topological defects with half-integer topological charge, which are only possible in systems with nematic symmetry. The activity of the colony stems from cell migration: \textit{M. xanthus} cells exert traction forces on the substrate to self-propel along their long axis. As cells move in the dense colony, they mechanically interact with neighboring cells. These mechanical interactions give rise to active stresses in the cell monolayer. The presence of active stresses is evidenced by the self-propulsion of $+1/2$ topological defects, which is a fingerprint of active nematic systems \cite{Doostmohammadi2018}. As they move, cells continuously slide past each other, rearranging and exchanging neighbors in the colony. Thus, the colony is in a fluid state. Altogether, we conclude that the bacterial colony forms an active nematic fluid layer. In the following, we present a minimal continuum model to describe the flows of cells in this bacterial monolayer.

\subsubsection{Nematic order} \label{nematic-order}

Based on the physics of liquid crystals \cite{deGennes-Prost}, we describe the orientational order of the cell monolayer in terms of the nematic order-parameter tensor $\bm{Q}$. In two dimensions, $Q_{\alpha\beta} = S \,[ 2\,\hat{n}_\alpha \hat{n}_\beta - \delta_{\alpha\beta}]$, where $S$ is the scalar strength of the order parameter, and $\hat{\bm{n}} = (\cos\theta,\sin\theta)$ is the unitary director field, with $\theta$ the orientation angle. The cartesian components of $\bm{Q}$ are therefore
\begin{subequations} \label{eq Q}
\begin{align}
Q_{xx} &= - Q_{yy} = S \cos(2\theta),\\
Q_{xy} &= Q_{yx} = S \sin(2\theta).
\end{align}
\end{subequations}
In terms of $\bm{Q}$, we write the nematic free energy as \cite{Beris1994,Selinger2016}
\begin{multline} \label{eq nematic-free-energy}
F = \int \left[\frac{a}{2} Q_{\alpha\beta} Q_{\alpha\beta} + \frac{b}{4} (Q_{\alpha\beta} Q_{\alpha\beta})^2\right. \\
\left.+ \frac{L}{2} (\partial_\alpha Q_{\beta\gamma})(\partial_\alpha Q_{\beta\gamma})\right]\,\dd^2\bm{r}.
\end{multline}
The first two terms come from a Landau expansion in the order parameter, with $a<0$ in the nematic phase, and $b>0$ for stability. These coefficients define the saturation value of the order parameter strength, $S_0=\sqrt{-a/(2b)}$. The last term corresponds to the Frank energy, which accounts for the orientational elasticity of liquid crystals. We have taken the usual one-constant approximation, where $L$ is the single orientational elastic modulus of the material, which is directly related to the Frank elastic constant $K$ \cite{deGennes-Prost,Beris1994,Selinger2016}.

For simplicity, we ignore flow-alignment couplings, which only result in a small correction to the self-propulsion speed of $+1/2$ defects \cite{Shankar2018b}. Moreover, we assume that the nematic relaxation rate is much faster than the typical strain rates associated with the flows. In this limit \cite{Giomi2014a,Shankar2018b}, the order-parameter field relaxes to the configuration that minimizes the free energy \cref{eq nematic-free-energy}, i.e.
\begin{equation} \label{eq nematic-equilibrium}
\frac{\delta F}{\delta Q_{\alpha\beta}} = \left(a + b\, Q_{\gamma\delta} Q_{\gamma\delta}\right) Q_{\alpha\beta} - L \bm{\nabla}^2 Q_{\alpha\beta} = 0.
\end{equation}
Given that $\bm{Q}$ is a rank-2 symmetric and traceless tensor, it can be described in terms of a single complex field $\psi = Q_{xx} + i Q_{xy} = S\, e^{2i\theta}$. In terms of this field, the equilibrium condition \cref{eq nematic-equilibrium} reads
\begin{equation} \label{eq psi}
L \bm{\nabla}^2 \psi = \left(a + 2b |\psi|^2\right) \psi,
\end{equation}
whose solution determines the nematic order in the system.

\subsubsection{Force balance} \label{force-balance}

We treat the cell monolayer as a thin layer of an incompressible fluid with an upper free boundary described by the height profile $h(\bm{r},t)$, where $\bm{r}$ is the two-dimensional position vector. The incompressibility condition imposes $\bm{\bm{\nabla}}\cdot\bm{u} = 0$ on the three-dimensional flow field $\bm{u}$. Choosing the $\hat{\bm{z}}$ axis perpendicular to the plane, and decomposing the velocity field into in-plane and out-of-plane components, $\bm{u}=(\bm{v},u_z)$, the incompressibility condition can be recast as
\begin{equation} \label{eq height}
\partial_t h(\bm{r},t) = - \bm{\bm{\nabla}}\cdot \int_0^{h(\bm{r},t)} \bm{v}(\bm{r},z,t) \,\dd z.
\end{equation}
This equation expresses that the divergence of the planar flow field gives rise to height changes of the fluid layer.

Following previous work \cite{Kawaguchi2017}, we initially neglect planar pressure gradients in the force balance. This corresponds to assuming that, under compression, cells in one layer do not build up pressure because they can readily move to another layer. This assumption will be relaxed in \cref{counterflow}. Moreover, we treat the cell colony as a dry system \cite{Ramaswamy2003,Marchetti2013}, i.e. we neglect internal viscous forces with respect to cell-substrate friction. This approximation is well justified based on the experimentally measured correlation length of the velocity field, $\ell_{\text{v}}= 3.9\pm0.1$ $\mu$m (\cref{velocity-correlation}). This correlation length is smaller than the nematic correlation length $\ell_{\text{n}}= 16.0\pm0.5$ $\mu$m (\cref{nematic-correlation}), and even smaller than the length of a single cell, $l\approx 7$ $\mu$m. Therefore, long-range hydrodynamic interactions can be safely neglected in this system. Furthermore, because we assume that the nematic order parameter rapidly relaxes to its equilibrium configuration (\cref{eq nematic-equilibrium}), both the antisymmetric and the flow-alignment contributions to the stress tensor vanish \cite{Julicher2018}. With these simplifications, the force balance reduces to a balance between frictional and active forces in the nematic cell monolayer:
\begin{equation} \label{eq force-balance}
\xi_{\alpha\beta} v_\beta = f^{\text{a}}_\alpha.
\end{equation}
Here, $f^{\text{a}}_\alpha = \partial_\beta \sigma^{\text{a}}_{\alpha\beta}$ is the active force density that arises from the anisotropic active stress $\sigma^{\text{a}}_{\alpha\beta} = -\zeta Q_{\alpha\beta}$ developed in the colony, with coefficient $\zeta>0$ for extensile and $\zeta<0$ for contractile stresses. Finally, $\xi_{\alpha\beta}$ is the friction coefficient matrix.

To account for the anisotropy of cell-substrate interactions, we take \cite{Kawaguchi2017}
\begin{equation} \label{eq anisotropic-friction}
\xi_{\alpha\beta} = \xi_0 \left[ \delta_{\alpha\beta} - \epsilon\, Q_{\alpha\beta}\right].
\end{equation}
Here, the first term accounts for isotropic friction with coefficient $\xi_0$, whereas the second term makes the friction coefficient depend on the local cell orientation via the nematic order parameter $Q_{\alpha\beta}$. Thus, $\epsilon$ is the friction anisotropy. For $\epsilon>0$, friction is stronger in the direction perpendicular to the alignment of cells. Even though, to some extent, friction anisotropy might stem from the elongated shape of \textit{M. xanthus} cells, its main source is likely the anisotropic response of cell-substrate focal adhesions. Previous experiments suggest that these adhesions barely yield in the direction perpendicular to the long cell axis, but are readily detached and reformed along the cell axis during gliding motility \cite{Balagam2014}. These observations suggest that $\epsilon>0$ for \textit{M. xanthus}.

\subsection{Nematic order and flow fields around topological defects} \label{defects}

In this section, we solve the model introduced in \cref{model} to obtain the nematic order and cell flow fields around isolated topological defects in the cell monolayer. We consider only the lowest-energy topological excitations, i.e. point defects with topological charge $q = \pm 1/2$.

\subsubsection{Nematic order} \label{nematic-order-defects}

For $\pm 1/2$ defects, the orientation angle is given by $\theta =  q \phi = \pm \phi/2$, where $\phi$ is the polar angle defined with respect to the defect's symmetry axis. Consequently, and assuming that $S(\bm{r}) = S(r)$, where $r$ is the radial distance from the defect core, \cref{eq psi} reduces to
\begin{equation} \label{eq S}
S''(r) + \frac{1}{r} S'(r) - \frac{1}{r^2} S(r) = \frac{1}{\ell^2} \left[\frac{S^2(r)}{S_0^2} - 1\right] S(r),
\end{equation}
where we have defined the nematic healing length $\ell = \sqrt{L/|a|}$. With the conditions $S(0)=0$ and $S(\infty)=S_0$, the solution to this nonlinear equation determines the radial profile of nematic order around a topological defect. The full solution cannot be obtained analytically, but it can be approximated by the following Pad\'{e} approximant \cite{Pismen2006,Pismen1999}:
\begin{equation} \label{eq Pade}
S(r) = S_0 f(r/\ell);\qquad f(x) \approx x\sqrt{\frac{0.34 + 0.07 x^2}{1+0.41 x^2 + 0.07 x^4}}.
\end{equation}
Note that the nematic order-parameter profile $S(r)$ is the same for both positive and negative defects, which are only distinguished by their orientation-angle profiles $\theta(\phi)=\pm \phi/2$.

\subsubsection{Flow field} \label{flow-field-defects}

Based on the solutions of the nematic order around topological defects obtained above, and using \cref{eq Q}, we obtain the active-force density $f^{\text{a}}_\alpha = -\zeta \partial_\beta Q_{\alpha\beta}$. For a $+1/2$ defect, with $\theta = \phi/2$, the Cartesian components of the active-force density are
\begin{subequations}
\begin{align}
f_x^{\text{a},+} &= -\zeta \left[S'(r) + \frac{S(r)}{r}\right],\\
f_y^{\text{a},+} &= 0.
\end{align}
\end{subequations}
Using these expressions, we can solve the force balance, \cref{eq force-balance}, to obtain the velocity field around a $+1/2$ defect:
\begin{subequations} \label{eq flow-field+1/2}
\begin{align}
v^+_x(r,\phi) &= -\frac{\zeta}{\xi_0} \left[S'(r) + \frac{S(r)}{r}\right] \frac{1 + \epsilon\, S(r)\cos\phi}{1-\epsilon^2 S^2(r)}, \label{eq velx+1/2}\\
v^+_y(r,\phi) & = -\frac{\zeta}{\xi_0} \left[S'(r) + \frac{S(r)}{r}\right] \frac{\epsilon\, S(r)\sin\phi}{1-\epsilon^2 S^2(r)}.
\end{align}
\end{subequations}

Respectively, for a $-1/2$ defect, with $\theta = -\phi/2$, the active-force density reads
\begin{subequations}
\begin{align}
f_x^{\text{a},-} &= -\zeta \left[S'(r) - \frac{S(r)}{r}\right]\cos(2\phi),\\
f_y^{\text{a},-} &= \zeta \left[S'(r) - \frac{S(r)}{r}\right]\sin(2\phi).
\end{align}
\end{subequations}
Solving \cref{eq force-balance}, the velocity field around a $-1/2$ defect is given by
\begin{subequations} \label{eq flow-field-1/2}
\begin{align}
& \begin{multlined}
v^-_x(r,\phi) = -\frac{\zeta}{\xi_0} \left[S'(r) - \frac{S(r)}{r}\right]\times\\
\frac{ [1 + \epsilon\, S(r)\cos\phi] \cos(2\phi) + \epsilon\, S(r) \sin\phi \sin(2\phi) }{1-\epsilon^2 S^2(r)}, \label{eq velx-1/2}
\end{multlined}\\
&v^-_y(r,\phi) = \frac{\zeta}{\xi_0} \left[S'(r) - \frac{S(r)}{r}\right] \frac{ [2\cos\phi - \epsilon\, S(r)] \sin\phi}{1-\epsilon^2 S^2(r)}.
\end{align}
\end{subequations}

\subsubsection{Inflow toward +1/2 defects and outflow from -1/2 defects} \label{net-flux}

\begin{figure}[tb]
\begin{center}
\includegraphics[width=0.8\columnwidth]{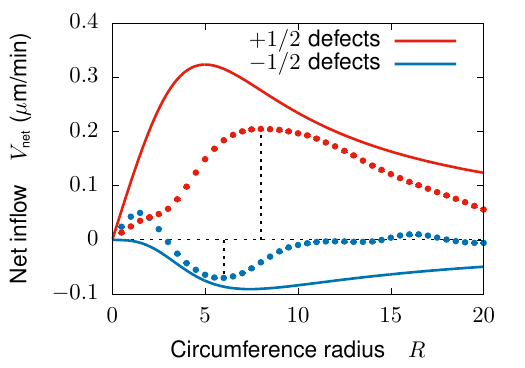}
\end{center}
\bfcaption{Net inflow around topological defects}{ $V_{\text{net}}$, defined in \cref{eq net-inflow}, is the average net inflow velocity through a circumference of radius $R$ centered at a topological defect. Points are experimental data obtained from the average flow fields in \cref{+1/2-flow-experiment,-1/2-flow-experiment}. Error bars (S.E.M.) are barely visible because they are smaller than the point size. Solid curves are the theoretical predictions given in \cref{eq net-inflow-prediction}, evaluated using the parameter values in \cref{t fits}. Vertical dashed lines indicate the radii at which the experimental net inflow magnitude is maximal. This maximal net inflow is presented in \cref{net-inflow}.} \label{net-inflow-radius}
\end{figure}

\begin{figure}
    \includegraphics[width=\columnwidth]{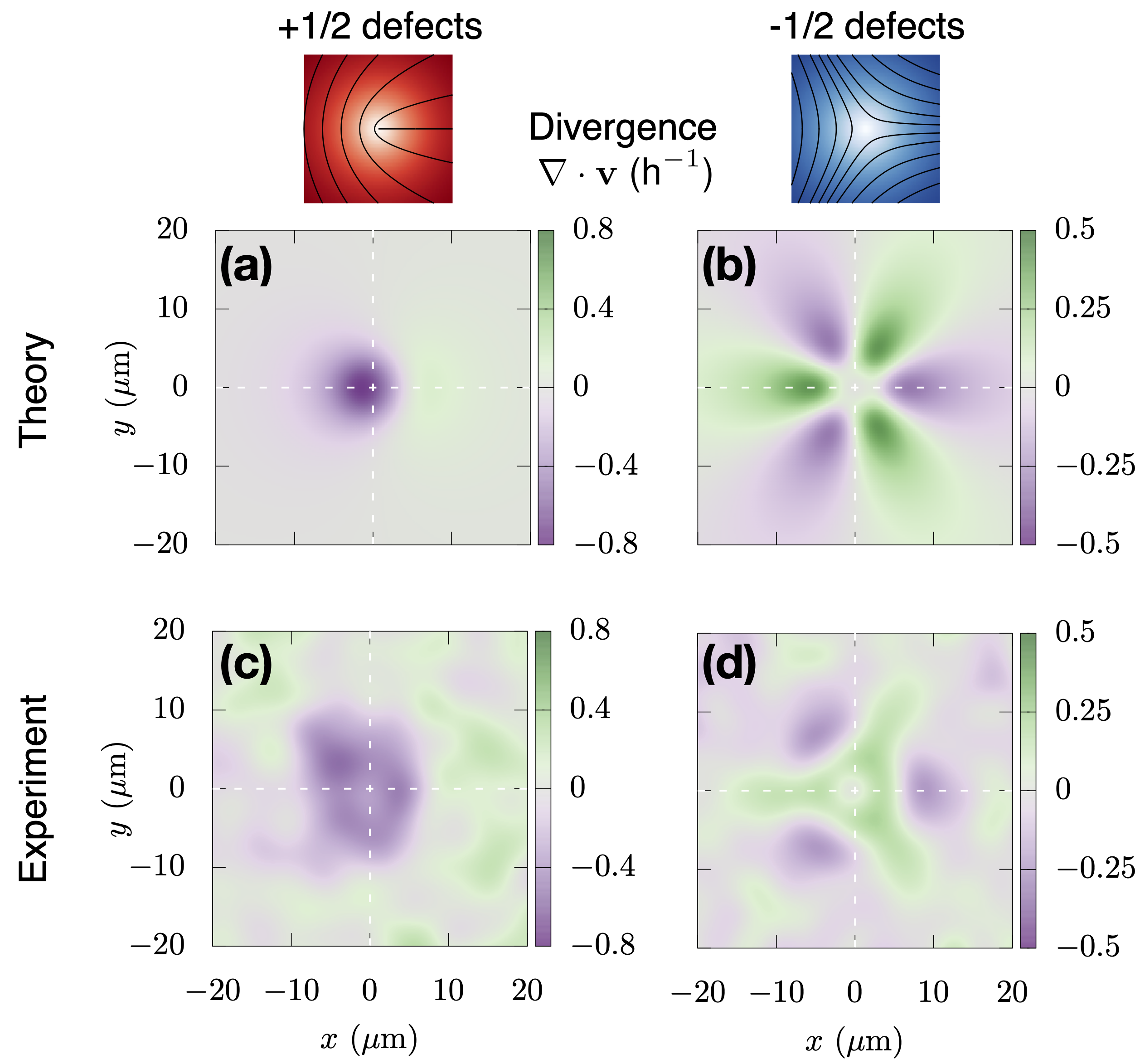}
  {\phantomsubcaption\label{+1/2-divergence-theory}}
  {\phantomsubcaption\label{-1/2-divergence-theory}}
  {\phantomsubcaption\label{+1/2-divergence-experiment}}
  {\phantomsubcaption\label{-1/2-divergence-experiment}}
    \bfcaption{Divergence of the flow field}{ Theoretically predicted (\subref*{+1/2-divergence-theory}, \subref*{-1/2-divergence-theory}) and experimentally measured (\subref*{+1/2-divergence-experiment}, \subref*{-1/2-divergence-experiment}) divergence fields around topological defects. The defect schematics show the order parameter (color map) and a few director-field lines. The divergence of the two-dimensional flow field, $\bm{\nabla}\cdot\bm{v}$, shows pronounced cell accumulation ($\bm{\nabla}\cdot\bm{v}<0$, purple) in front of $+1/2$ defects, and cell depletion ($\bm{\nabla}\cdot\bm{v}<0$, green) in three lobes along the axes of symmetry of $-1/2$ defects. The parameter values used to plot panels \subref*{+1/2-divergence-theory} and \subref*{-1/2-divergence-theory} are listed in \cref{t fits}.}
    \label{divergence}
\end{figure}

These velocity fields predict a net inflow toward $+1/2$ defects and a net outflow from $-1/2$ defects. Specifically, the number of cells $N(R)$ inside a circle $\mathcal{C}$ of radius $R$ centered at a defect changes as
\begin{equation}
\frac{\dd N(R)}{\dd t} = - \oint_{\partial\mathcal{C}} \bm{J}\cdot \dd\bm{s} = - \rho_0 R \int_0^{2\pi} v_r(R,\phi) \,\dd\phi.
\end{equation}
Here, we used that $\bm{J} = \rho_0 \bm{v}$ is the cell flux, with $\rho_0$ the uniform monolayer cell density, and $\dd\bm{s} = \hat{\bm{r}}(\phi)R\,\dd\phi$ is normal to the circle boundary. Thus, we define the average net flow into the circle of radius $R$ as
\begin{equation} \label{eq net-inflow}
V_{\text{net}} (R) = -\frac{1}{2\pi}\int_0^{2\pi} v_r(R,\phi)\,\dd\phi.
\end{equation}
Using \cref{eq flow-field+1/2,eq flow-field-1/2} in $v_r = v_x \cos\phi + v_y\sin\phi$, we obtain
\begin{equation}
v_r^+ (r,\phi) = -\frac{\zeta}{\xi_0} \left[S'(r) + \frac{S(r)}{r}\right] \frac{\cos\phi + \epsilon\, S(r)}{1-\epsilon^2 S^2(r)},
\end{equation}
\begin{equation}
v_r^- (r,\phi) = -\frac{\zeta}{\xi_0} \left[S'(r) - \frac{S(r)}{r}\right] \frac{\cos(3\phi) + \epsilon\, S(r)}{1-\epsilon^2 S^2(r)},
\end{equation}
from which we obtain
\begin{equation} \label{eq net-inflow-prediction}
V_{\text{net}}^{\pm}(R) = \frac{\zeta}{\xi_0} \left[S'(R) \pm \frac{S(R)}{R}\right] \frac{2\pi \epsilon\, S(R)}{1-\epsilon^2 S^2(R)}.
\end{equation}
This quantity quantifies the asymmetry of the flow field around defects due to friction anisotropy $\epsilon$. For $\epsilon>0$, as in \textit{M. xanthus}, there is a net inflow toward $+1/2$ defects, $V_{\text{net}}^+ (R)>0$, and a net outflow from $-1/2$ defects, $V_{\text{net}}^- (R)<0$ (\cref{net-inflow-radius}). The net inflow/outflow toward/from positive/negative defects is also apparent in the divergence of the velocity field (\cref{divergence}).

\subsubsection{Fits to experimental data} \label{fits}

To compare our theoretical predictions to the measured flow fields, we fitted the predicted velocity profile along the midline of the defects, i.e. the symmetry line that defines the $\hat{\bm{x}}$ axis. Along this axis, the velocity field is entirely longitudinal, $\bm{v}(x,0) = v_x(x,0) \,\hat{\bm{x}}$, with $v_y(x,0)=0$. Moreover, along the midline, polar coordinates become $r=|x|$ and $\phi = 0,\pi$ for the positive and negative sides of the $\hat{\bm{x}}$ axis. Using these relations, we fit \cref{eq velx+1/2,eq velx-1/2} to velocity profiles along the midline of $+1/2$ and $-1/2$ defects, respectively. \Cref{eq velx+1/2,eq velx-1/2} depend on four parameters: the saturation value $S_0$ of the nematic order parameter, the healing length $\ell$, the ratio $\zeta/\xi_0$ between the active stress and the isotropic friction coefficients, and the friction anisotropy $\epsilon$. The first two parameters characterize the nematic order around defects (\cref{eq Pade}), whereas the last two parameters are related to the force balance \cref{eq force-balance,eq anisotropic-friction}. To perform the fits, we fixed the value of $S_0=0.992\pm 0.003$, as given by experimental measurements (\hyperref[methods]{Methods}), and we left $\ell$, $\zeta/\xi_0$, and $\epsilon$ as fitting parameters. The optimal parameter values are listed in \cref{t fits}, and the fits are shown in \cref{midline-flow-fits} (red and blue curves for $+1/2$ and $-1/2$ defects, respectively).

Whereas the values of the healing length $\ell$ for $+1/2$ and $-1/2$ defects are compatible, the values of $\zeta/\xi_0$ and $\epsilon$ are not. The difference in parameter values for $+1/2$ and $-1/2$ defects suggests that cells experience a different effective friction around each defect type. This difference may stem from the different mechanical details of the layer formation and hole opening processes, for example due to the surface tension of the water layer that wets the colony, which are not accounted for in our minimal model. Therefore, explaining the difference in the parameter values that we obtain from the fits requires a more detailed model that we defer to future work.

\begin{table}[tb]
\begin{center}
\begin{tabular}{lcc}
Parameter&$+1/2$ defects&$-1/2$ defects\\\hline
$\ell$ ($\mu$m)&$2.5\pm 0.1$&$2.54\pm 0.02$\\
$\zeta/\xi_0$ ($\mu$m$^2$/min)&$1.77\pm0.09$&$4.65\pm 0.04$\\
$\epsilon$&$0.80\pm 0.01$&$0.214\pm0.006$
\end{tabular}
\caption{Optimal values of parameters obtained by separately fitting the velocity profile along the midline of $+1/2$ and $-1/2$ defects (\cref{midline-flow-fits}).} \label{t fits}
\end{center}
\end{table}

\begin{figure}[tbh]
\begin{center}
\includegraphics[width=0.8\columnwidth]{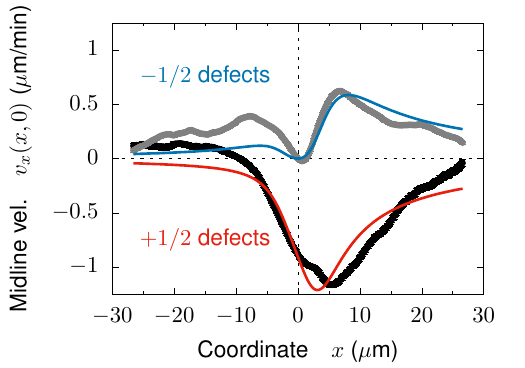}
\end{center}
\bfcaption{Simultaneous fits of the velocity profiles of $+1/2$ and $-1/2$ defects}{ The black (grey) data points and the red (blue) curve are the experimentally measured and the fitted midline velocity profile of $+1/2$ ($-1/2$) defects. These simultaneous fits to $+1/2$ and $-1/2$ defects, with the common set of parameter values given in \cref{t simultaneous-fits}, are poorer than the separate fits to $+1/2$ and $-1/2$ defects shown in \cref{midline-flow-fits}.} \label{simultaneous-fits}
\end{figure}

\begin{table}[tbh!]
\begin{center}
\begin{tabular}{lc}
Parameter&$+1/2$ and $-1/2$ defects\\\hline
$\ell$ ($\mu$m)&$2.55\pm 0.07$\\
$\zeta/\xi_0$ ($\mu$m$^2$/min)&$2.09\pm0.07$\\
$\epsilon$&$0.77\pm 0.01$
\end{tabular}
\caption{Optimal parameter values obtained by simultaneously fitting the velocity profile along the midline of $+1/2$ and $-1/2$ defects (\cref{simultaneous-fits}).} \label{t simultaneous-fits}
\end{center}
\end{table}

Finally, if instead of fitting the velocity profiles of $+1/2$ and $-1/2$ defect separately, we fit them simultaneously, we obtain poorer fits (\cref{simultaneous-fits}). The common set of optimal parameter values obtained from these fits is given in \cref{t simultaneous-fits}.

\subsection{Compressibility effects: explanation of the counterflow in front of $+1/2$ defects} \label{counterflow}

$+1/2$ defects exhibit an overall active motion along their axis. This defect motion corresponds to the negative velocity ($v_x<0$) that we measure along most of the defect axis (\cref{midline-flow-fits}). However, in the front of $+1/2$ defects, we measure a cell flow toward the defect core, opposite to the overall motion of the defect. This frontal counterflow corresponds to the measured positive velocity ($v_x>0$) in the region $x \lesssim -10$ $\mu$m (\cref{midline-flow-fits}). This counterflow cannot be explained by the model in \cref{model,defects}. That model predicts that the active force is negative $f^{\text{a},+}_x <0$ at every point around an isolated $+1/2$ defect. As a consequence, the velocity is also negative everywhere, and there is no counterflow.

In this section, we propose a possible explanation for the measured counterflow. We argue that the counterflow may stem from pressure gradients that we neglected in the force balance \cref{eq force-balance} based on the assumption that, under compression, cells are readily extruded from the cell monolayer. If cells do not readily move to an upper layer, the non-uniform motion of a $+1/2$ defect generates bulk deformations in the cell monolayer. In particular, the front of the defect experiences compression, leading to a pressure increase. Therefore, a pressure gradient is established along the defect axis. This expectation is consistent with the pressure gradient measured along the axis of $+1/2$ defects in epithelial monolayers \cite{Saw2017}. This pressure gradient, in turn, generates a flow toward the lower pressures at the defect core, i.e. opposite to the active flow responsible for defect motion. Far from the defect core, where active forces are small, the pressure-driven counterflow may overcome the active flow, thus possibly explaining our experimental observation of counterflow. In the following, we minimally extend our model to implement these ideas and predict the counterflow.

\subsubsection{Active nematic model with a non-uniform cell density} \label{non-uniform-density}

In \cref{model,defects}, we assumed that the cell density in the basal cell monolayer was uniform, $\rho(\bm{r}) = \rho_0$. As a consequence, the pressure in the basal layer was also uniform. Hence, pressure gradients vanished, and the force balance reduced to \cref{eq force-balance},
\begin{equation}
\xi_{\alpha\beta} v^0_\beta = f_\alpha^{\text{a}},
\end{equation}
where $f_\alpha^{\text{a}}$ is the active-force density, and $\bm{v}^0$ is the flow field corresponding to that uniform-density situation. This flow field, given by \cref{eq flow-field+1/2}, captured most features of the experimentally measured flow fields around topological defects. However, it did not capture the counterflow measured in front of the $+1/2$ defects (\cref{midline-flow-fits}).

To capture the counterflow, we assume that cells do not readily go into an upper layer in response to non-uniform flows. In this case, pressure gradients build up due to density inhomogeneities in the basal layer. Therefore, force balance becomes
\begin{equation} \label{eq force-balance-pressure}
\xi_{\alpha\beta} v_\beta = -\partial_\alpha P + f_\alpha^{\text{a}},
\end{equation}
where $P$ is the pressure in the basal cell layer. We assume that this pressure depends on the basal cell density through a linear equation of state:
\begin{equation} \label{eq state}
P(\rho) = B \frac{\rho - \rho_0}{\rho_0} + P_0,
\end{equation}
where $B$ is the bulk modulus of the monolayer, and $P_0 = P(\rho_0)$ is the pressure of the uniform-density reference state. Introducing \cref{eq state} into \cref{eq force-balance-pressure}, we obtain
\begin{equation} \label{eq complete-force-balance}
\xi_{\alpha\beta} v_\beta = - \frac{B}{\rho_0}\partial_\alpha \rho  + f_\alpha^{\text{a}}.
\end{equation}
Both the friction coefficient $\xi_0$ and the active stress coefficient $\zeta$ are per unit area (see \cref{force-balance}). Therefore, not only the pressure but also these coefficients could depend on cell density. Hereafter, in order to show that pressure gradients alone are sufficient to produce a counterflow, we ignore these possible additional dependencies. Moreover, given that the flow field obtained from the uniform-density approximation was close to the experimental measurements, we propose to explain the counterflow based on small perturbations around the uniform-density solution: $\rho(\bm{r}) = \rho_0 + \delta\rho(\bm{r})$ and $\bm{v}(\bm{r}) = \bm{v}^0(\bm{r}) + \delta\bm{v}(\bm{r})$. For these perturbations, and with the aforementioned assumptions, the force balance \cref{eq complete-force-balance} implies
\begin{equation} \label{eq flow-perturbations}
\xi_{\alpha\beta} \delta v_\beta = - \frac{B}{\rho_0}\partial_\alpha \delta\rho.
\end{equation}

To obtain the flows induced by pressure gradients from \cref{eq flow-perturbations}, we need to specify the basal cell density field $\rho(\bm{r})$. This field obeys the continuity equation
\begin{equation}
\frac{\partial \rho(\bm{r},t)}{\partial t} = - \bm{\nabla}\cdot\left[\rho(\bm{r},t) \bm{v}(\bm{r},t)\right] - k_{\text{extr}}(\bm{r},t),
\end{equation}
where $k_{\text{extr}}(\bm{r},t)$ is the rate of cell extrusion from the basal monolayer to upper cell layers. If cells readily extrude upon compression, the steady-state extrusion rate is
\begin{equation} \label{eq extrusion-rate}
k_{\text{extr}}(\bm{r}) = - \bm{\nabla} \cdot \left[\rho(\bm{r}) \bm{v}(\bm{r})\right].
\end{equation}
In the uniform-density situation of \cref{model,defects}, the extrusion rate reduces to
\begin{equation}
k_{\text{extr}}^0(\bm{r}) = -\rho_0 \bm{\nabla}\cdot \bm{v}^0(\bm{r}).
\end{equation}
More generally, when we introduce the small perturbations $\rho(\bm{r}) = \rho_0 + \delta\rho(\bm{r})$ and $\bm{v}(\bm{r}) = \bm{v}^0(\bm{r}) + \delta\bm{v}(\bm{r})$, \cref{eq extrusion-rate} expands into
\begin{multline} \label{eq extrusion-rate-expansion}
k_{\text{extr}}(\bm{r}) \approx -\rho_0 \bm{\nabla}\cdot \bm{v}^0(\bm{r}) - \rho_0 \bm{\nabla}\cdot\delta\bm{v}(\bm{r})\\
 - \delta\rho(\bm{r}) \bm{\nabla}\cdot \bm{v}^0(\bm{r}) - \bm{v}^0(\bm{r})\cdot \bm{\nabla}\delta\rho(\bm{r}).
\end{multline}
This expression determines the extrusion rate if cells readily move to a new layer under compression. However, if cells do not readily move to a new layer, the rate of cell extrusion is no longer given by \cref{eq extrusion-rate-expansion}. Rather, some of the terms on the right-hand side of \cref{eq extrusion-rate-expansion} correspond to in-plane density inhomogeneities that do not lead to cell extrusion. Here, we assume that the extrusion rate is given by the first two terms of \cref{eq extrusion-rate-expansion}, which are proportional to the uniform density $\rho_0$: $k_{\text{extr}}(\bm{r}) = -\rho_0 \bm{\nabla}\cdot \bm{v}(\bm{r})$. Consequently, the remaining two terms of \cref{eq extrusion-rate-expansion} must balance:
\begin{equation} \label{eq density-perturbations}
\delta\rho(\bm{r}) \bm{\nabla}\cdot \bm{v}^0(\bm{r}) + \bm{v}^0(\bm{r})\cdot \bm{\nabla}\delta\rho(\bm{r}) = 0.
\end{equation}
\Cref{eq density-perturbations} is a closed differential equation for the in-plane density perturbation $\delta\rho(\bm{r})$. Its solution can then be inserted into \cref{eq flow-perturbations} to obtain the flow perturbations $\delta\bm{v}(\bm{r})$ that may account for the counterflow.

\subsubsection{Cell density and velocity perturbations along the defect axis} \label{density-defects}

\begin{table}[tb]
\begin{center}
\begin{tabular}{lc}
Parameter&$+1/2$ defects\\\hline
$\ell$ ($\mu$m)&$3.1\pm 0.1$\\
$\zeta/\xi_0$ ($\mu$m$^2$/min)&$2.47\pm 0.09$\\
$\epsilon$&$0.63\pm 0.02$\\
$\beta$ ($\mu$m$^3$/min$^2$)&$-0.32\pm 0.04$
\end{tabular}
\caption{Optimal values of parameters obtained by fitting the velocity profile along the defect midline, taking into account density inhomogeneities due to monolayer compression (\cref{midline-flow-fits}, yellow curve). The parameter $\beta\equiv BJ/(\rho_0\xi_0)$ is related to the compressibility $B$ of the cell monolayer and to the cell flux $J<0$ induced by the associated pressure gradient.} \label{t fits-compressible}
\end{center}
\end{table}

\begin{figure}[tbh!]
\begin{center}
\includegraphics[width=0.8\columnwidth]{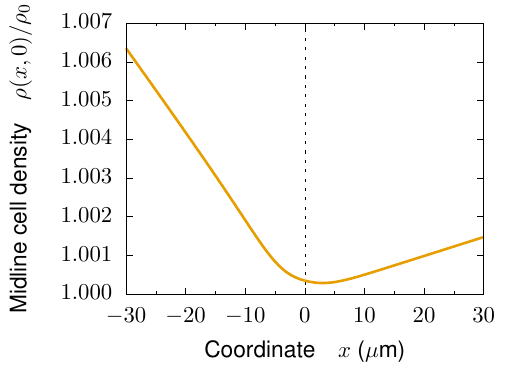}
\end{center}
\bfcaption{Predicted cell density profile along the midline of $+1/2$ defects}{ The total cell density is $\rho(x,0) = \rho_0 + \delta\rho(x,0)$, where $\delta\rho(x,0)$ is given by \cref{eq density-perturbation-solution}, with $v_x^0(x,0)$ obtained from the fits (parameter values in \cref{t fits-compressible}), and $J=10^{-4}$ ($\mu$m$\cdot$min)$^{-1}$.} \label{density-perturbation}
\end{figure}

Here, we obtain an approximate solution of \cref{eq density-perturbations} along the midline of a $+1/2$ defect. We approximate the divergence of the flow along the midline by its longitudinal component only, i.e. $\bm{\nabla}\cdot\bm{v}^0|_{y=0} \approx \dd v^0_x/\dd x|_{y=0}$. Moreover, since the flow field along the midline is purely longitudinal, $\bm{v}^0 (x,0) = v^0_x(x,0)\, \hat{\bm{x}}$, \cref{eq density-perturbations} reduces to
\begin{equation}
\delta\rho(x,0)\left.\frac{\dd v^0_x}{\dd x}\right|_{y=0} + v^0_x(x,0) \left.\frac{\dd\delta\rho}{\dd x}\right|_{y=0} \approx 0.
\end{equation}
This equation can be recast as
\begin{equation}
\frac{\dd}{\dd x}\left(\left.(\delta\rho\,v^0_x) \right|_{y=0}\right) = 0.
\end{equation}
The solution to this equation is
\begin{equation} \label{eq density-perturbation-solution}
\delta\rho(x,0) = \frac{J}{v^0_x(x,0)},
\end{equation}
where $J$ is an undetermined integration constant, which corresponds to a uniform cell flux along the defect midline. Along the midline, the force balance for the velocity perturbation, \cref{eq flow-perturbations}, reads
\begin{equation}
\xi_{xx}(x,0) \delta v_x(x,0) = - \frac{B}{\rho_0} \frac{\dd \delta\rho(x,0)}{\dd x}.
\end{equation}
Using \cref{eq density-perturbation-solution}, the velocity perturbation can be finally written entirely in terms of the unperturbed velocity field $v_x^0$ (given in \cref{eq flow-field+1/2}):
\begin{equation} \label{eq velocity-perturbation-midline}
\delta v_x(x,0) = \frac{B J}{\rho_0 \xi_0} \frac{1}{1-\epsilon S(r) \cos\phi} \frac{1}{(v^0_x(x,0))^2}\frac{\dd v^0_x (x,0)}{\dd x}.
\end{equation}
Along the midline, polar coordinates become $r=|x|$ and $\phi = 0,\pi$ for the positive and negative sides of the $\hat{\bm{x}}$ axis.

\subsubsection{Fit to experimental data}

Finally, we fit the total velocity profile along the midline,
\begin{equation}
v_x(x,0) = v^0_x(x,0) + \delta v_x(x,0),
\end{equation}
to the experimental data. Here, $v_x^0(x,0)$ is obtained from \cref{eq velx+1/2} and $\delta v_x(x,0)$ is given by \cref{eq velocity-perturbation-midline}. In addition to the original parameters $\ell$, $\zeta/\xi_0$, and $\epsilon$, now there is an additional fitting parameter $\beta\equiv BJ/(\rho_0\xi_0)$ characterizing the compressibility of the cell monolayer and the cell flux due to the associated pressure gradient. The result of the fit is shown in \cref{midline-flow-fits} (yellow curve), and the optimal values of the parameters are listed in \cref{t fits-compressible}. The fact that $\beta<0$ reflects that $J<0$, meaning that the cell density increases as the negative $v_x^0(x,0)$ decreases in magnitude when moving away from the defect core (\cref{density-perturbation}).

\end{document}